\documentclass[aps,prb,showpacs,a4paper,twocolumn,superscriptaddress,groupedaddress]{revtex4}  % for review and submission
\usepackage{graphicx}
\usepackage[toc,page]{appendix}
\usepackage{multirow}
\usepackage{dcolumn}   % needed for some tables
\usepackage{bm}        % for math
\usepackage{amssymb}   % for math
\usepackage{amsmath}
\usepackage{natbib}
\usepackage[bookmarks=false]{hyperref}
\newcommand\scalemath[2]{\scalebox{#1}{\mbox{\ensuremath{\displaystyle #2}}}}

\begin{document}
%\preprint{Version 10; 25 July 2005}
% the following line is for submission, including submission to the arXiv!!
%\hspace{5.2in} \mbox{Fermilab-Pub-04/xxx-E}

\title{Model of coherent optical spin manipulation through hot trion states in p-doped InAs/GaAs quantum dots}
%\title{Model of the enhanced circular dichroism in the excited states emission from p-doped InAs/GaAs quantum dots}
\author{G. Slavcheva}
\affiliation{Blackett Laboratory, Imperial College London,\\
Prince Consort Road, London SW7
2AZ, United Kingdom}
\affiliation{Mediterranean Institute of Fundamental Physics, Via Appia Nuova 31, 00040 Rome, Italy}
\email{slavcheva@mifp.eu}
%\affiliation{$^1$ Blackett Laboratory, Imperial College London,\\
%Prince Consort Road, London SW7
%2AZ, United Kingdom,\\$^2$ Mediterranean Institute of Fundamental Physics, Via Appia Nuova 31, 00040 Rome, Italy}
%\email{slavcheva@mifp.eu}

%\date{\today}

\begin{abstract}
A new generalised group-theoretical approach, based on master Maxwell-pseudospin equations, is proposed to explain recently observed enhanced circular dichroism in the excited state emission from p-doped quantum dot ensembles under resonant circularly polarised excitation into hot trion states, herein referred to as "spin-filtering effect". The theory agrees remarkably well with polarised time-resolved photoluminescence experiments, yielding largely unknown inter- and intra-shell spin relaxation time scales. This approach allows to predict optimum pulse parameters for control of spin dynamics, which will enable exploitation of the effect in all-optical spin-based quantum technologies.

\end{abstract}
%PACS categories:
% 78.20.Bh: Theory, models and numerical simulations
% 75.40.Mg: Numerical simulation studies
% 78.47.jd: Time resolved luminescence
% 78.47.da: Excited states
% 78.67.Hc: Quantum dots
% 71.35.Pq: Charged excitons (trions)
% 71.70.Gm: Exchange interactions
% 78.67.-n: Optical properties of low-dimensional and nanoscale materials and structures
% 72.25.Fe: Optical creation of spin polarized carriers
% 72.25.Rb: Spin relaxation and scattering
% 42.50.Ct: Quantum description of interaction of light and matter and related
\maketitle

\section{Introduction}
The ability to exercise strict control over the quantum dot (QD) emission properties is paramount for applications of QDs in quantum technologies and requires in-depth understanding of their energy level structure. Three-dimensional confinement of carriers in charged quantum dots eliminates many bulk spin relaxation mechanisms but others, such as the exchange and hyperfine interactions, are enhanced. Optical circularly polarised excitation of charged QDs creates excited (hot) charged exciton (trion) states with a complex fine structure, due to exchange interactions \cite{Kavokin}. Exchange interactions play an important role in lifting the level degeneracy and determine the energy-level configuration and splitting. The fine structure and the entailed optical selection rules and spin relaxation mechanisms become critical for successful implementation of QD-based optical spin manipulation schemes, based on excited trion transitions \cite{Carter}$^,$\cite{Kim&Economou}. On the other hand, the use of radiative cascades from QD excited states is a principal method for generation of quantum entanglement for applications of QDs as single-photon emitters and polarisation-entangled sources. The exchange interaction and the fine structure splitting of the energy levels involved in radiative cascades from QD excited states are fundamental as they determine the polarisation and entanglement of the emitted photons \cite{Poem}. The fine energy level structure is revealed in the excited charged excitonic spectra. For these reasons interest in the excited charged excitons is growing. However, unlike the ground trion 'singlet' negative ($X^-$) [5-8] and positive ($X^+$)[9-11] QD state that has been relatively well studied and understood, there have been very few reports on the negative hot trion, $X^{-*}$,  in n-doped [12-18], and even fewer on the hot positive trion, $X^{+*}$, in p-doped QDs [19-21].

Optical manipulation of a single electron (hole) spin, confined to a QD, through the resonantly driven charged exciton (trion) ground singlet transition, is currently considered as one of the most promising schemes for implementation of spin-based quantum computing, due to extended spin lifetimes, limited by the hyperfine interaction [22-25]. The requirement of a resonantly driven ground trion transition in an inhomogeneously broadened ensemble of charged QDs, however, represents an obstacle for the scalability of architectures based on this quantum system.

Here I show that the limitation of resonant excitation of the ground trion singlet state can be overcome by an optical excitation into the excited trion states, taking advantage of a recently discovered effect, herein referred to as "spin-filtering effect". The latter consists of an enhanced photo-induced circular dichroism in the excited state emission of p-doped QDs, relative to the ground state, under non-resonant circularly polarised optical excitation \cite{Taylor}. The degree of polarisation is nearly doubled when resonant excitation into an excited dot level is used, due to increased spin injection efficiency. In addition, non-monotonic dependence of the degree of spin polarisation on the optical pulse power is observed, allowing to maximise it by optimising the pulse characteristics. This spin-filtering effect is promising for realisation of high-fidelity schemes for all-optical spin manipulation, since the increased polarisation degree contrast for the two pulse helicities enables highly efficient selective excitation and readout of the spin-up and spin-down populations.

In this work I theoretically demonstrate the possibility of using the hot trion states in p-doped QDs for efficient spin manipulation by short circularly polarised optical pulses. In Sec. II I give details of the polarised TRPL experiment and develop a comprehensive theory and a quantitative dynamical model of the optically induced circular dichroism, observed in the time-resolved photoluminescence (TRPL) from ensembles of p-doped QDs under pulsed optical excitation. I show that the enhanced photo-induced circular dichroism of the excited state emission, compared to the ground state, stems from the specific trion fine structure, comprising two sets of degenerate states, dynamically coupled through spin flip-flop processes, which induce asymmetry in the allowed optical transitions for left- and right-circularly polarised light. In Sec. III I summarize the results from the numerical simulations carried out on a realistic single QD layer structure and compare the calculated polarised TRPL traces with the experimental ones directly in the time domain. In particular, I discuss the dependence of the experimental data on detector type and the importance of convolving the simulation data with the detector response function for a proper comparison with the experimental data. The experimental data are fitted by the model, thereby yielding largely unknown intra- and inter-shell spin relaxation times, summarised in Table 1, and an estimate of the magnitude of the effect is provided. The concluding section IV summarizes the main results and outlines future model applications.

\section{Theoretical model}
\begin{figure}
%\vspace{-10pt}
%\resizebox{\columnwidth}{!}{\includegraphics{Fig1.eps}}
\includegraphics[height=5 cm, width=6 cm]{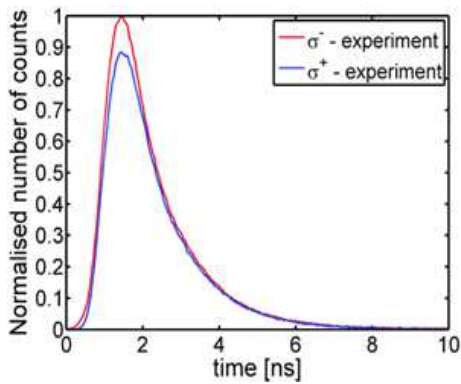}
\vspace{-10pt}
\caption[Fig 1] {\label{fig:Fig 1}(Color online) Experimentally detected polarised TRPL traces for $\sigma^-$ and $\sigma^+$ excitation.}\label{fig:experiment_PL}
\end{figure}

The QD samples employed in this work are grown by MBE on a semi-insulating GaAs $(001)$ substrate \cite{Harbord}. The QD layers are sandwiched between two GaAs barriers and the dot areal density is $~2\times10^{10} \,\mathrm{cm^{-2}}$ with an average uncapped height $h=4\,\mathrm{nm}$ ($N_a=5\times10^{22} \,\mathrm{m^{-3}}$). PL experiments under resonant left- ($\sigma^-$) or right- ($\sigma^+$) circularly polarised excitation into excited dot states with pumping wavelength $\lambda_{res}=1065 \,\,\mathrm{nm}$ were performed on QD ensembles, nominally doped with one hole. The polarised PL is detected at the ground singlet $X^{+}$ transition, exhibiting a peak at $\lambda_{det}=1148 \,\,\mathrm{nm}$. The TRPL experiments are carried out with short ($\tau=50 \,\mathrm{ps}$) optical pulses and the PL decay in time is given in Fig. \ref{fig:experiment_PL}, clearly showing marked difference in the spin polarisation degree for both pulse helicities.
%%%%%%%%%%%%%%%%%%%%%%%%%%%%%%%%%%%%%%%%%%%%%%%%%%%%%%%%%%%%%%%%%%%%%%%%%%%%%%%
%%%%%%%%%%%%%%%%%%%%%%%%%%%%%%%%%%%%%%%%%%%%%%%%%%%%%%%%%%%%%%%%%%%%%%%%%%%%%%%%%%%%%

\begin{figure}
\resizebox{\columnwidth}{!}{\includegraphics {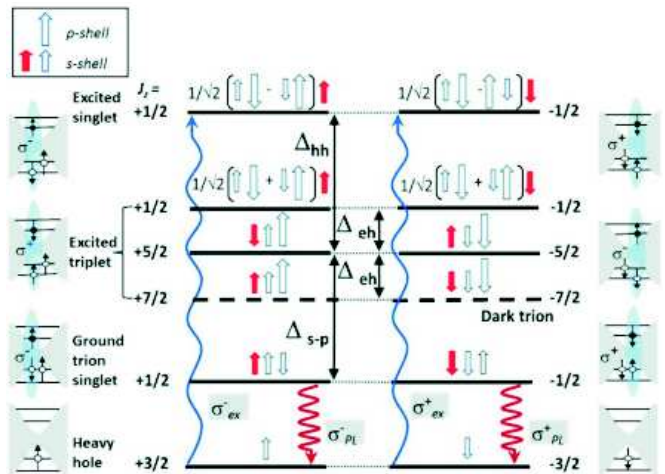}}
\vspace{-10pt}
\vspace{-10pt}
\caption[Fig 2] {(Color online) Energy levels and spin configurations of the hot $X^{+*}$ trion states under $\sigma^-$ (left panel) and $\sigma^+$ (right panel) optical excitation (upward blue wavy arrows). Small solid (red) and open (blue) arrowheads represent s-shell electron and hole total spin projection $J_z$; large open (blue) arrow -- p-shell hole spin. Downward dark red wavy arrows denote the PL; shaded blue ellipse: exciton coupling in the radiative states. The excited singlet is split from the triplet by $\Delta_{hh}$; the excited triplets are split by $\Delta_{eh}$. $\Delta_{s-p}$--energy separation between p-shell and s-shell trion states.}\label{fig:trion_energy_level_scheme} \label{fig:hot_trion_energy_level_scheme}
\end{figure}

To understand the observed complex polarisation dynamics, we consider quasi-resonant $\sigma^-$ or $\sigma^+$ circularly polarised excitation of a p-doped QD ensemble into the p-shell in the presence of a resident s-shell hole, and subsequent cascade relaxation to the bright trion ground (singlet) state, whose decay is detected as PL. The optical excitation creates $X^{+*}$ states, consisting of one electron-hole pair in the s-shell and a resident hole in the p-shell ($1e^11h^12h^1$), grouped in four degenerate doublets \cite{Kavokin}. The energy-level diagram and the spin configurations of the $X^{+*}$ trion is given in Fig.~\ref{fig:trion_energy_level_scheme}.
%%%%%%%%%%%%%%%%%%%%%%%%%%%%%%%%%%%%%%%%%%%%%%
%%%%%%%%%%%%%%%%%%%%%%%%%%%%%%%%%%%%%%%%%%%%%%%%%%
The isotropic hole-hole ($h-h$) exchange interactions give rise to the singlet-triplet splitting, $\Delta_{hh}\sim 1-10 \,\mathrm{meV}$. The isotropic part of the electron-hole ($e-h$) exchange interaction splits the triplet states into three equally spaced levels with spacing $\Delta_{eh}\sim 0.1 - 1\,\mathrm{meV}$. The anisotropic $e-h$ exchange interaction (AEI) leads to mixing of these states.

The system Hamiltonian of a circularly polarised pulse, resonantly coupled to an ensemble of 6-level resonant absorbers with density $N_a$, is given by:
%\begin{widetext}
\vspace{-5pt}
\setlength{\mathindent}{-5pt}
\begin{equation}
\hat H^\mp=
\end{equation}
\begin{equation*}
%\begin{multline}
\label{eq:Hamiltonian}
%\hat H^\mp=
%\\
\hbar\left( \scalemath{0.55}{ {\begin{array}{*{20}c}
   0 & 0 & 0 & 0 & 0 & { - \frac{1}{2}\left( {\Omega _x  \mp i\Omega _y } \right)}  \\
   0 & {\omega _0  - \Delta _{hh}  - \Delta _{sp} } & 0 & 0 & 0 & 0  \\
   0 & 0 & {\omega _0  - \Delta _{hh}  - \Delta _{eh} } & 0 & 0 & 0  \\
   0 & 0 & 0 & {\omega _0  - \Delta _{hh} } & 0 & 0  \\
   0 & 0 & 0 & 0 & {\omega _0  - \Delta _{hh}  + \Delta _{eh} } & 0  \\
   { - \frac{1}{2}\left( {\Omega _x  \pm i\Omega _y } \right)} & 0 & 0 & 0 & 0 & {\omega _0 }  \\
\end{array}}}\right)
\end{equation*}
%\end{multline}
%\end{widetext}
where $\mp$ correspond to $\sigma^-$ ($\sigma^+$) polarisation, $\Omega _x  = \wp \frac{{E_x }}{\hbar }$, $\Omega _y  = \wp
\frac{{E_y }}{\hbar }$ are the time-dependent Rabi-frequencies
associated with $E_x$ and $E_y$ electric field components and $\omega_0$ and $ \wp$
are the resonant transition frequency and the optical dipole matrix element of the ground to excited singlet state ($|1\rangle\rightarrow|6\rangle$) transition.

\begin{figure}
\vspace{-10pt}
%\includegraphics[scale=0.3]{Fig3.eps}
%\vspace{-10pt}
\resizebox{\columnwidth}{!}{\includegraphics{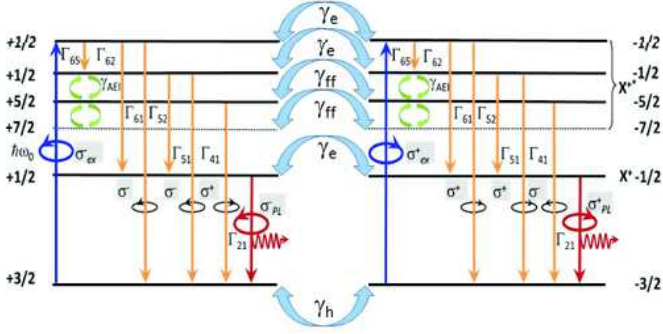}}
\vspace{-20pt}
\caption[Fig 3] {(Color online) Discrete-level model of the $X^{+*}$ states. The two degenerate six-level systems are coupled via spin decoherence mechanisms: $\gamma_h$ -- hole-spin decoherence, $\gamma_e$ -- electron-spin decoherence, $\gamma_{ff}$ -- spin decoherence due to spin flip-flop processes (AEI). Circularly polarised transitions are designated by $\circlearrowleft$ or $\circlearrowright$. Upward blue arrows: $\sigma^-_{ex}$ and $\sigma^+_{ex}$ resonant pumping, downward orange arrows: radiative or nonradiative transitions; Dark red arrows: detected polarised PL. Curved green arrows: spin flip-flop coupling due to AEI. }\label{fig:energy_level_scheme_spin_relaxation}
\end{figure}

Spin decoherence mechanisms that need to be considered include $e-h$ AEI, which remains relevant for excited trion states, and the hyperfine interaction between nuclei and either electrons \cite{Braun} or holes through dipole-dipole interaction \cite{Eble}. The two degenerate 6-level systems, for $\sigma^-$ and $\sigma^+$-excitation, are coupled through transverse spin decoherence mechanisms, as shown in Fig.~\ref{fig:energy_level_scheme_spin_relaxation}.

%%%%%%%%%%%%%%%%%%%%%%%%%%%%%%%%%%%%%%%%%%%%%%%%%%
\setlength{\mathindent}{0pt}
%%%%%%%%%%%%%%%%%%%%%%%%%%%%%%%%%%%%%%%%%%%%%%%%%%%%%%%%
The time evolution of the 6-level quantum system in the presence of relaxation processes (Fig.~\ref {fig:energy_level_scheme_spin_relaxation}), under time-dependent external perturbation (laser pulse) is governed by a master equation \cite{Slavcheva} for the $N^2-1=35$-dimensional real state (pseudospin) vector, $S_j$:
\vspace{-5pt}
\begin{eqnarray}
\begin{aligned}
\label{eq:pseudospin}
& \frac{{\partial S_j }}{{\partial t}} = \left\{
\begin{array}{l}
 f_{jkl} \gamma _k S_l  + \frac{1}{2}Tr\left( {\hat \sigma \,\hat \lambda _j }
 \right) - \frac{1}{{T_j }}\left( {S_j  - S_{je} } \right),\,\,\,\, \\
 \,\,\,\,\,\,\,\,\,\,\,\,\,\,\,\,\,\,\,\,\,\,\,\,\,\,\,\,\,\,\,\,\,\,\,\,\,\,\,\,\,
 \,\,\,\,\,\,\,\,\,\,\,\,\,\,\,\,\,\,\,\,\,\,\,\,\,\scalemath{0.8}{j = 1,2,...,30} \\
 f_{jkl} \gamma _k S_l  + \frac{1}{2}Tr\left( {\hat \sigma \,\hat \lambda _j } \right),\,\,\,\scalemath{0.8}{j = 31,32,..,35} \\
 \end{array} \right.
 \end{aligned}
\end{eqnarray}
where $\lambda_j$ are the generators, $\it{f}$ is the fully antisymmetric tensor of the structure constants of SU(6) group and $\mathbf{\gamma}$ is the torque vector given in Appendix A, Eqs. (A5), (A6) and (A8), respectively.
$T_{j}$ are phenomenological non-uniform transverse spin decoherence times describing the relaxation of $S_{j}$ toward their equilibrium values $S_{je}$. The longitudinal spin relaxation is incorporated in the matrix $\hat \sigma=diag(Tr(\hat \Gamma_i \hat \rho))$, where $\hat \rho$ is the density matrix and $\hat \Gamma_i$,$i=1,...,6$, are spin relaxation rate matrices, given explicitly in Appendix B, Eq. (B2) for $\sigma^-$ and Eq. (B10) for $\sigma^+$ excitation.

\begin{figure}
%\resizebox{\columnwidth}{!}{\includegraphics {Fig4.eps}}
\includegraphics[scale=0.4]{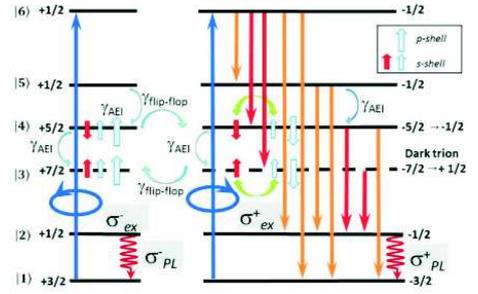}
\vspace{-10pt}
\caption[Fig 4] {(Color online) Energy-level diagram for $\sigma^+$ excitation, modified by the spin flip-flop processes in the triplet states (level $|3\rangle$ and $|4\rangle$, right panel). Downward (orange) arrows denote radiative and non-radiative transitions before spin-flip flop. The green double-headed arrows represent the spin flip-flop of the electron and the hole, changing the total spin of state $|3\rangle$ from $|-7/2\rangle \rightarrow |+1/2\rangle$, and of state $|4\rangle$ from $|-5/2\rangle \rightarrow |-1/2\rangle$. This results in additional allowed transitions, denoted by red arrows.}\label{fig:sigma_plus_level_scheme}
\end{figure}

The excited triplet states in the two degenerate systems, where the photo-generated and the resident hole spins are parallel (Fig.~\ref{fig:hot_trion_energy_level_scheme}), are coupled via spin  flip-flop processes, whereby the electron and the hole flip simultaneously their spin (level $|3\rangle$ and $|4\rangle$ in Fig.~\ref{fig:energy_level_scheme_spin_relaxation}). This $e-h$ AEI-mediated process plays a significant role in n-doped and p-doped QD spin dynamics \cite{Braun}. Due to dynamical coupling of the two degenerate systems through spin flip-flop processes and simultaneous electron and hole spin reversal, the total spin projection of these states changes and as a result additional transitions become allowed due to spin selection rules. For instance, the total spin projection of state $|3\rangle$ changes from $|-7/2\rangle \rightarrow |+1/2\rangle$, and of state $|4\rangle$ from $|-5/2\rangle \rightarrow |-1/2\rangle$, thus making possible the transitions $|6\rangle\rightarrow|4\rangle$, $|6\rangle\rightarrow|3\rangle$, $|4\rangle\rightarrow|2\rangle$ and $|3\rangle\rightarrow|2\rangle$, denoted by red arrows in Fig.~\ref{fig:sigma_plus_level_scheme}.
Therefore more transitions at any one time become allowed in one polarisation (here without loss of generality chosen as $\sigma^+$) than in the other ($\sigma^-$).

%%%%%%%%%%%%%%%%%%%%%%%%%%%%%%%%%%%%%%%%%%%%%%%%%%%%%%%%%%%%%%%%%%%%%%%%%%%%%%%%%%%%%%%%%%%%%%
\begin{figure}
%\vspace{-20pt}
\resizebox{\columnwidth}{!}{\includegraphics{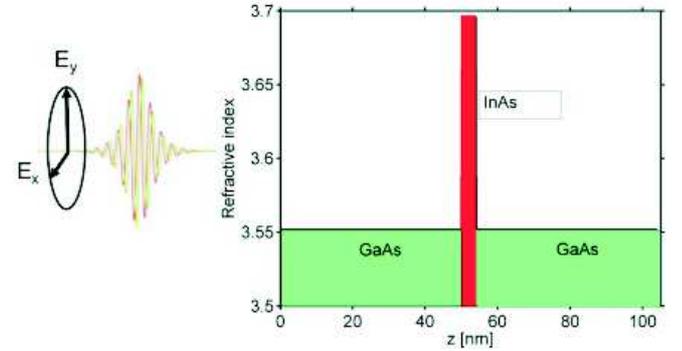}}
\vspace{-10pt}
\caption[Fig 5] {(Color online) Refractive index profile of the simulation domain, containing InAs QD layer ($h=4 \,\mathrm{nm}$), embedded between two GaAs barrier regions, $50 \,\mathrm{nm}$ each. The circularly polarised pulse, modelled by two x- and y- linearly polarised waves, phase shifted by $\pi/2$, and its E-field components, are schematically shown.}\label{fig:simulation domain}
\end{figure}

%%%%%%%%%%%%%%%%%%%%%%%%%%%%%%%%%%%%%%%%%%%%%%%%%%%%%%%%%%%%%%%%%%%%%%%%%%%%%%%%%%%%
The simulations domain consists of an InAs QD layer with nominal thickness given by the height of the typical QD ($h=4\,\mathrm{nm}$), embedded between two GaAs barrier regions each with thickness $50 \,\mathrm{nm}$ (see Fig. ~\ref{fig:simulation domain}).
%%%%%%%%%%%%%%%%%%%%%%%%%%%%%%%%%%%%%%%%%%%%%%%%%%%%%%%%%%%%%%%%%%%%%%%%%%%%%%%%%
%%%%%%%%%%%%%%%%%%%%%%%%%%%%%%%%%%%%%%%%%%%%%%%%%%%%%%%%%%%%%%%%%%%%%%%%%%%%%%%%%%%%

Maxwell's curl equations for a circularly polarised laser pulse propagating along the QD growth direction, chosen as $z$-axis read:
%\begin{subequations}
\vspace{-5pt}
\begin{equation}
\label{eq:Maxwell}
\begin{aligned}
& \frac{\partial H_x \left({z,t}\right)}{\partial t} = \frac{1}{\mu }\frac{{\partial E_y \left({z,t}\right)}}{{\partial z}} \\
& \frac{{\partial H_y \left( {z,t} \right)}}{{\partial t}} =  - \frac{1}{\mu }\frac{{\partial E_x \left( {z,t} \right)}}{{\partial z}} \\
& \frac{{\partial E_x \left( {z,t} \right)}}{{\partial t}} =  - \frac{1}{\varepsilon }\frac{{\partial H_y \left( {z,t} \right)}}{{\partial z}} - \frac{1}{\varepsilon }\frac{{\partial P_x \left( {z,t} \right)}}{{\partial t}} \\
& \frac{{\partial E_y \left( {z,t} \right)}}{{\partial t}} =
\frac{1}{\varepsilon }\frac{{\partial H_x \left( {z,t}
\right)}}{{\partial z}} - \frac{1}{\varepsilon }\frac{{\partial P_y
\left( {z,t} \right)}}{{\partial t}}
\end{aligned}
\end{equation}
%\end{subequations}
where $P_x$ and $P_y$ are the macroscopic polarisation vector components. Eqs.(\ref{eq:pseudospin}) and (\ref{eq:Maxwell}) are coupled through the macroscopic polarisation induced in the medium by the electromagnetic wave (see Appendix C):
\begin{equation}
\label{eq:polarisation}
\begin{array}{l}
 P_x  =  - \wp N_a S_5  \\
 P_y  =  \mp \wp N_a S_{20}  \\
 \end{array}
\end{equation}
where $\mp$ corresponds to $\sigma^-$ $(\sigma^+)$ excitation.

The circularly polarized Gaussian pulse at the left boundary of the simulation domain, $z=0$, is modelled by two x- and y-linearly polarised waves with resonant carrier frequency, $\omega_0$, phase shifted by $\pi/2$:
%\begin{eqnarray}
\begin{equation}
\begin{array}{l}
\label{eq:initialpulse} \sigma ^ -  \left\{ \begin{array}{l}
 E_x \left( {z = 0,t} \right) = E_0 e^{ - {{\left( {t - t_0 } \right)^2 } \mathord{\left/
 {\vphantom {{\left( {t - t_0 } \right)^2 } {t_d^2 }}} \right.
 \kern-\nulldelimiterspace} {t_d^2 }}}\cos (\omega _0 t) \\
 E_y \left( {z = 0,t} \right) =  - E_0 e^{ - {{\left( {t - t_0 } \right)^2 } \mathord{\left/
 {\vphantom {{\left( {t - t_0 } \right)^2 } {t_d^2 }}} \right.
 \kern-\nulldelimiterspace} {t_d^2 }}}\sin (\omega _0 t) \\
 \end{array} \right.\\
 \\
\sigma ^ +  \left\{ \begin{array}{l}
 E_x \left( {z = 0,t} \right) = E_0 e^{ - {{\left( {t - t_0 } \right)^2 } \mathord{\left/
 {\vphantom {{\left( {t - t_0 } \right)^2 } {t_d^2 }}} \right.
 \kern-\nulldelimiterspace} {t_d^2 }}}\cos (\omega _0 t) \\
 E_y \left( {z = 0,t} \right) = E_0 e^{ - {{\left( {t - t_0 } \right)^2 } \mathord{\left/
 {\vphantom {{\left( {t - t_0 } \right)^2 } {t_d^2 }}} \right.
 \kern-\nulldelimiterspace} {t_d^2 }}}\sin (\omega _0 t) \\
 \end{array} \right.
 \end{array}
 \end{equation}
%\end{eqnarray}
where $E_{0}$ is the initial field amplitude. The system of equations (\ref{eq:pseudospin}-\ref{eq:initialpulse}) is discretised in space and time ( $\Delta z=1 \,\mathrm{\AA}$, $\Delta t=3.33 \times 10 ^{-4} \,\mathrm{fs}$) and solved numerically directly in the time domain using the Finite-Difference Time-Domain (FDTD) technique.

\section{Numerical simulations and comparison with experiment}
\subsection{Simulated dynamics}
An optical pulse with pulse duration $\tau=50\,\mathrm{ps}$ (selected to match the experiment) is injected at $\textit{z}=0$. The pulse center frequency  $\omega_{0}$, is tuned in resonance with the energy splitting between the ground heavy-hole level $|1\rangle$ and the excited singlet trion state $|6\rangle$: $\Delta E_{1 \to 6}  = E_{X^+}  + \Delta _{s - p}  + \Delta _{hh}$, where the ground trion singlet energy $E_{X^+}= 1.0815 \,\mathrm{eV}$ is determined from the resonant PL spectra at $\lambda_{det}$. $\Delta _{s - p}\sim 73\,\,\mathrm{meV}\,\,$ is inferred from the PL spectra \cite{Taylor} and $\Delta_{hh}=12 \,\mathrm{meV}$, $\Delta_{eh}\sim0.5 \,\mathrm{meV}$ are taken in agreement with \cite{Kavokin}$^,$ \cite{Warming}.

The pulse area below the pulse envelope is chosen to be $\pi$, corresponding to $E_0= 2.69\times10^5\,\mathrm{Vm^{-1}}$ (Eq.(\ref{eq:initialpulse})), assuming dipole matrix element $\wp=9.83\times10^{-29} \,\mathrm{Cm}$. The pulse power, corresponding to a $\pi-$pulse is equivalent to $754 \,\mathrm{mW}$ ($\sim$ half of the peak pulse power delivered by the Ti:sapphire laser). The dipole moment of the $X^+$ ground singlet transition, $\wp\sim (7\pm2)\times 10^{-29}$, [30-32] is taken as initial approximation. Radiative decay rate $\Gamma_{21}\sim 1.27 \,\,\mathrm{ns^{-1}}$ of the $X^+$ ground singlet transition is measured in \cite{Taylor}. Spontaneous emission rates $\Gamma_{41}\sim 1.35 \,\mathrm{ns^{-1}}$, $\Gamma_{51}\sim \Gamma_{61} \sim 1.2 \,\mathrm{ns^{-1}}$ are estimated, using the Fermi golden rule \cite{Yariv}.

Non-radiative spin relaxation times $\sim 1-7 \,\mathrm{ps^{-1}}$ are calculated in \cite{Narvaez}, although conflicting values for the $p\rightarrow s$ intershell relaxation times can be found in the literature: from fast, $1.5 - 9 \,\mathrm{ps^{-1}}$ to very slow $750 \,\mathrm{ps^{-1}} - 7.7 \,\,\mathrm{ns^{-1}}$. Our simulations show that slow intershell relaxation times do not reproduce the observed TRPL time evolution, therefore we assume $\Gamma_{52}\sim 5 \,\mathrm{ps^{-1}}$. The intrashell ($s*\rightarrow s$) rate $\Gamma_{62}\sim 5 \,\mathrm{ps^{-1}}$ is taken in agreement with \cite{Zibik}. The timescale of the intershell ($s* \rightarrow p$) hole spin relaxation is largely unknown. Our simulations show that these processes occur on a tens of ps timescale, but faster than $25 \,\mathrm{ps^{-1}}$ rates lead to unphysical negative spin population. We assume $\Gamma_{65}\sim 50 \,\,\mathrm{ps^{-1}}$, correctly reproducing the experimental TRPL trace.

The spin decoherence rates for the transverse relaxation processes (Fig.~\ref{fig:energy_level_scheme_spin_relaxation}) are taken $\gamma_e=500 \,\mathrm{ps^{-1}}$ for electron spin decoherence through hyperfine interaction with the lattice ion spins\cite{Braun}, $\gamma_h=14 \,\mathrm{ns^{-1}}$ for the hole spin decoherence due to dipole-dipole interaction \cite{Eble}; the spin-flip rate, $\gamma_{ff}=\gamma_{AEI} \sim 125 \,\mathrm{ps^{-1}}$ (AEI $\sim$ a few tens of $\mathrm{\mu eV}$ \cite{Braun}).

For $\sigma^-$ excitation, the dot density, $N_a$, and the dipole moment matrix element,$\wp$, have been identified as key parameters significantly affecting the time trace. The non-radiative decay times, $\tau_{62}$ and $\tau_{52}$, are critical for obtaining the correct amplitude. For $\sigma^+$ excitation variation of the radiative decay times, $\tau_{51}$ and $\tau_{41}$, significantly reshapes the TRPL trace.

The set of parameters producing best fit to the experimental data is given in Table I.
\begin{table*}
%\begin{table}
  \caption[Parameters]{Characteristic parameters used in the simulations of the TRPL following a $T_p=50 \,\mathrm{ps}$ pulse excitation with $\sigma^-$ and $\sigma^+$ polarised light. The decay timescales between a pair of levels $|i\rangle$ and $|j\rangle$ are denoted by $\tau_{ij}$, and $\tau_{AEI}$ is the decay timescale associated with the anisotropic e-h exchange interaction; meaning of all parameters explained in the text.}
\label{tab:bestfit-parameters}
\begin{center}
%    \begin{tabular}{l*{6}{c}r}
    \begin{tabular}{ | c | c | c | c | c | c | c | c | c | c | c |}
    \hline
     \multirow{2}{*}{Parameter}
     & $\tau_{21}$ & $\tau_{32}$ & $\tau_{41}$ & $\tau_{42}$ & $\tau_{51}$ & $\tau_{52}$ & $\tau_{61}$ & $\tau_{62}$ & $\tau_{63}$ & $\tau_{64}$ \\
     & $\,\mathrm{[ns]}$ & $\,\mathrm{[ns]}$ & & $\,\mathrm{[ns]}$ & & $\,\mathrm{[ps]}$ & $\,\mathrm{[ns]}$ & $\,\mathrm{[ps]}$ & $\,\mathrm{[ps]}$ & $\,\mathrm{[ns]}$
     \\
     \hline
    $\sigma^-$ & 1.27 & & 1.35 ns &  & 1.2 ns & 5 & 1.2 & 5 &  &   \\ \hline
    $\sigma^+$ & 1.27 & 1 ns & 10 ps & 1.5 & 7.5 ps & 500 & 1.2 & 10 & 800 & 1  \\ \hline
    \hline
    \multirow{2}{*}{Parameter}
    & $\tau_{65}$ & $\tau_{AEI}$ & $\lambda_{res}$ & $E_{X^+}$ &
    $\Delta_{sp}$ & $\Delta_{hh}$ & $\Delta_{eh}$ &
    $\wp$ & $N_a$ & $E_{0}$ \\
    & $\,\mathrm{[ps]}$ & $\,\mathrm{[ps]}$ & $\,\mathrm{[nm]}$ & $\,\mathrm{[eV]}$ & $\,\mathrm{[meV]}$ & $\,\mathrm{[meV]}$ & $\,\mathrm{[meV]}$ & $\,\mathrm{[Cm]}$ & $\,\mathrm{[m^{-3}]}$ & $\,\mathrm{[V m^{-1}]}$
    \\
    \hline
    $\sigma^-$ & 50 & 125 & 1065 & 1.0815 & 73 & 12 & 0.5 & $9.83\times10^{-29}$ & $5\times10^{22}$ & $2.69\times10^{5}$ \\ \hline
    $\sigma^+$ & 50 & 125 & 1065 & 1.0815 & 73 & 12 & 0.5 & $9.83\times10^{-29}$ & $5\times10^{22}$ & $2.69\times10^{5}$ \\ \hline
    \end{tabular}
\end{center}
%\end{table}
\end{table*}
%%%%%%%%%%%%%%%%%%%%%%%%%%%%%%%%%%%%%%%%%%%%%%%%%%%%%%%%%%%%%%%%%%%%%%%%%%%%%%%%%

\begin{figure}
   %\begin{tabular}{c}
   \vspace{-10pt}
   \resizebox{\columnwidth}{!}{\includegraphics{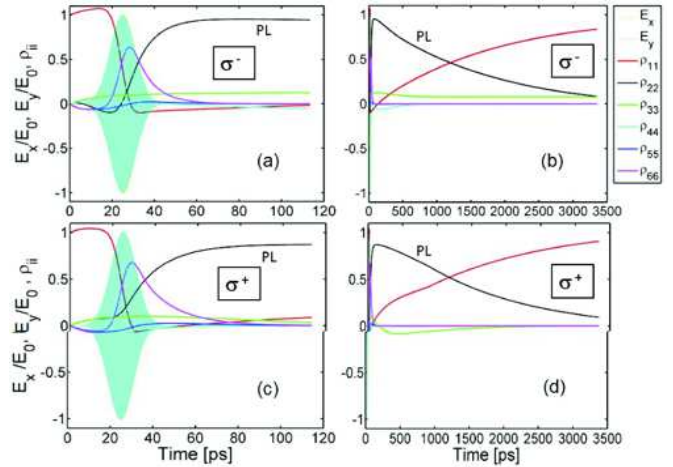}}
   \vspace{-20pt}
   %\includegraphics[scale=0.3]{Fig6.eps}
   %\vspace{-10pt}
   \caption[Fig 6]
   %>>>> use \label inside caption to get Fig. number with \ref{}
   { \label{fig:Fig 6}
    (Color online) Simulated short-time (a snapshot at $t= 120 \,\,\mathrm{ps}$ -- left panel) and long-time (a snapshot at $t= 3.5 \,\,\mathrm{ns}$ -- right panel) spin population dynamics of all six levels, following  $\sigma^-$ (a, b) and $\sigma^+$ (c, d) circularly polarised Gaussian $\pi$-pulse with pulse duration $T_p=50 \,\mathrm{ps}$ and E-field components $E_x$ and $E_y$. The spin population of level $|2\rangle$, $\rho_{22}$ (black curve), is proportional to the detected PL emission.}\label{fig:time evolution}
\end{figure}

\begin{figure}
%\vspace{-30pt}
%\includegraphics [scale=0.3]{Fig7.eps}
%\resizebox{\columnwidth}{!}{\includegraphics{Fig7.eps}}
%\includegraphics[height=6 cm, width=9 cm]{Fig7.eps}
\includegraphics[scale=0.3]{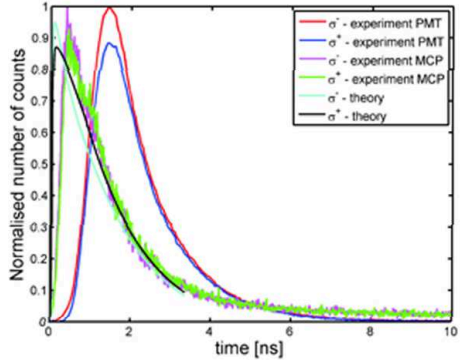}
%\vspace{-10pt}
\caption[Fig 7] {\label{fig:Fig 7}(Color online) Circular polarisation dynamics of the PL, following resonant excitation into p-shell states, experimentally detected from the QD $X^+$ ground singlet trion state by PMT and micro-channel plate (MCP) detectors, and theoretically computed polarised TRPL trace for $\sigma^-$ and $\sigma^+$ excitation ($\rho_{22}$ in Fig.~\ref{fig:time evolution}))}\label{fig:theory and detectors}
\end{figure}

The simulated spin population dynamics at a point in the QD layer using the best fit parameters (Table I), is plotted in Fig.~\ref{fig:time evolution}, showing significantly different long-time $\sigma^-$ and $\sigma^+$ TRPL decays ($\rho_{22}$ b, d).
%%%%%%%%%%%%%%%%%%%%%%%%%%%%%%%%%%%%%%%%%%%%%%%%%%%%%%%%%%%%
%%%%%%%%%%%%%%%%%%%%%%%%%%%%%%%%%%%%%%%%%%%%%%%%%%%%%%%%%%%%%
For comparison with the TRPL data, the computed TRPL time traces ($\rho_{22}$ in Fig.~\ref{fig:time evolution}) are plotted along with the experimentally detected photoluminescence by two detectors: InP/InGaAsP photomultiplier(PMT) and a microchannel plate detector (MCP) with extended S1 photocathode in Fig.~\ref{fig:theory and detectors}.
%%%%%%%%%%%%%%%%%%%%%%%%%%%%%%%%%%%%%%%%%%%%%%%%%%%%%%%%%%%%%%%%%%%%%%%%%%%%%%
%%%%%%%%%%%%%%%%%%%%%%%%%%%%%%%%%%%%%%%%%%%%%%%%%%%%%%%%%%%%%%%%%%%%%%%%%%%%%%%
It becomes apparent that the detected TRPL trace shape strongly depends on detector characteristics: the better the detector is, the closer it is to the simulated curve.

In order to properly compare our theory with the experimental data, we convolve our simulations with the photomultiplier detector response function (Figs. \ref{fig:detector_response_function}, \ref{fig:convolution}, Appendix D).

\subsection{Simulations length in time and dot density influence on the TRPL trace shape}
\begin{figure*}
%\vspace{20pt}
%\includegraphics[height=10 cm, width=16 cm]{Fig8.eps}
%\resizebox{\columnwidth}{!}{\includegraphics{Fig8.eps}}
\includegraphics[scale=0.6]{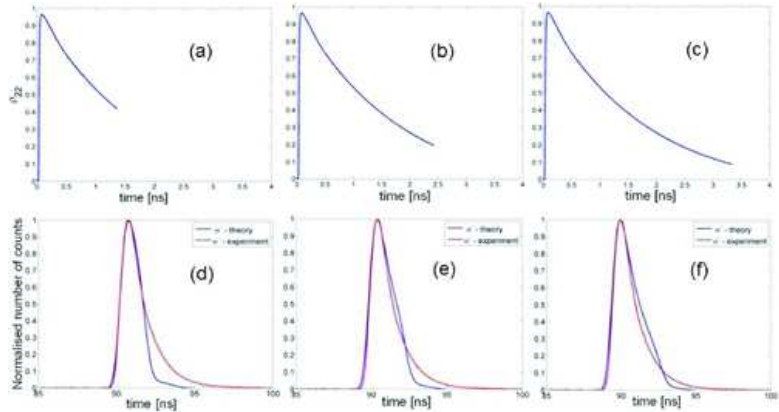}
%\vspace{-10pt}
\caption[Fig 8] {\label{fig:simulation_length}(Color online) Time evolution of the spin population of level $|2\rangle$, $\rho_{22}$, following $\sigma^-$ pulse excitation, for three different simulation lengths in time (a) $1.36 \,\mathrm{ns}$; (b) $1.95 \,\mathrm{ns}$; (c) $3.34 \,\mathrm{ns}$ (upper row) and the corresponding comparison between theory (after convolution of the raw simulation data, plotted on the upper row, with the detector response function from Fig. 4) (blue curve) and experimental TRPL trace (red curve) (lower row). The simulation parameters are as follows:$\tau_{21}=1.27 \,\mathrm{ns},\tau_{41}=1.35 \,\mathrm{ns},\tau_{51}=1.2 \,\mathrm{ns},\tau_{52}=5 \,\mathrm{ps},\tau_{61}=1.2 \,\mathrm{ns},\tau_{62}=5 \,\mathrm{ps},\tau_{65}=50 \,\mathrm{ps},\tau_{AEI}=125 \,\mathrm{ps},\lambda_{res}=1127 \,\,\mathrm{nm}, E_{X^+}=1.08 \,\mathrm{eV}, \wp_{1\rightarrow6}=8\times10^{-29}, N_a=5\times10^{22} \,\mathrm{m^{-3}}, E_0=3.3 \times10^5 \,\mathrm{Vm^{-1}}$}
\end{figure*}

The calculated TRPL shape, and therefore the agreement between the theory and experiment across the entire PL decay time, is very sensitive to the length in time of the simulations. In Fig. \ref{fig:simulation_length}, the spin population, $\rho_{22}$, time evolution for three different simulation lengths in time and the corresponding comparison between theory (after convolution with the detector response function) and experiment for a $\sigma^-$ pulse excitation are shown.
%%%%%%%%%%%%%%%%%%%%%%%%%%%%%%%%%%%%%%%%%%%%%%%%%%%%%%%%%%%
%%%%%%%%%%%%%%%%%%%%%%%%%%%%%%%%%%%%%%%%%%%%%%%%%%%%%%%%
To the best of our knowledge there have been no attempts to look at the TRPL decay beyond $2.5 \,\mathrm{ns}$\cite{Cortez}$^,$\cite{Braun}$^,$ \cite{Adler}. We show for the first time that the entire duration of the TRPL signal is crucial for correct reproduction of the PL time decay shape.

\begin{figure*}
%\vspace{20pt}
%\includegraphics[height=10 cm, width=14 cm]{Fig9.eps}
%\resizebox{\columnwidth}{!}{\includegraphics{Fig9.eps}}
\includegraphics[scale=0.7]{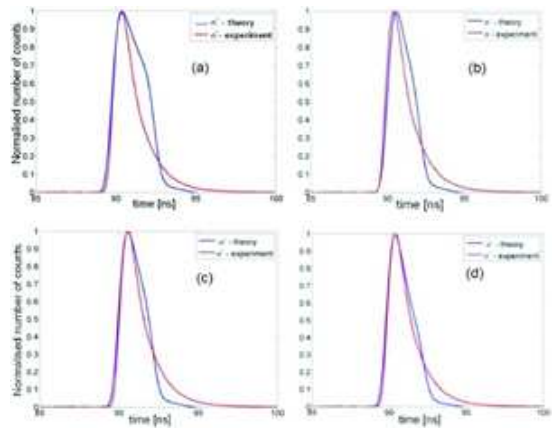}
%\vspace{-10pt}
\caption[Fig 9] {\label{fig:dot_density}(Color online) Comparison of the simulation data after convolution (blue line) with experimentally detected TRPL time trace (red line) for $\sigma^-$ pulse excitation with pulse duration $T_p=50 \,\mathrm{ps}$ and dot density (a) $N_a=5\times 10^{19} \,\mathrm{m^{-3}}$; (b) $N_a=1.5\times10^{22} \,\mathrm{m^{-3}}$; (c) $N_a=3\times10^{22} \,\mathrm{m^{-3}}$; (d) $N_a=5\times10^{22} \,\mathrm{m^{-3}}$, with the rest of simulation parameters as in Fig. \ref{fig:simulation_length}.}
\end{figure*}

The influence of the dot density on the TRPL time evolution trace is given in Fig. \ref{fig:dot_density} for four simulations with the rest of parameters kept fixed.
%%%%%%%%%%%%%%%%%%%%%%%%%%%%%%%%%%%%%%%%%
%%%%%%%%%%%%%%%%%%%%%%%%%%%%%%%%%%%%%%%%%%%%
We start with an estimate of the resonantly excited dot density on the order of $(0.2-0.3)N_a$ Fig. \ref{fig:dot_density}(a) and increase it stepwise to the density per unit volume of $5\times 10^{22} \,\mathrm{m^{-3}}$, calculated on the basis of the areal dot density of $2\times10^{10} \,\mathrm{cm^{-2}}$ and average height, $h=4 \,\mathrm{nm}$, of the QD layer. The departure of the theoretical curve from the experimental one, after the roll-over in Fig. \ref{fig:dot_density}(a) is quite substantial, and in order to save computational time, we interrupt the simulations as soon as the general trend is apparent (the kink in the theoretical curve is due to earlier interruption of the simulations). As the dot density increases by nearly three orders of magnitude (Fig. \ref{fig:dot_density}(b)), the discrepancy becomes smaller and the agreement between theory and experiment is significantly improved in Fig. \ref{fig:dot_density}(c) and (d). The best agreement between theory and experiment is achieved for the calculated volume density $N_a=5\times10^{22} \,\mathrm{m^{-3}}$. Further increase of the dot density to $N_a=7\times10^{22} \,\mathrm{m^{-3}}$ and beyond, does not lead to any improvement of the fit.
%\end{section}

\subsection{Comparison with experiment}
\begin{figure*}
\vspace{-10pt}
%\resizebox{\columnwidth}{!}{\includegraphics{Fig10.eps}}
%\includegraphics[height=12 cm, width=16 cm]{Fig10.eps}
\includegraphics[scale=0.5]{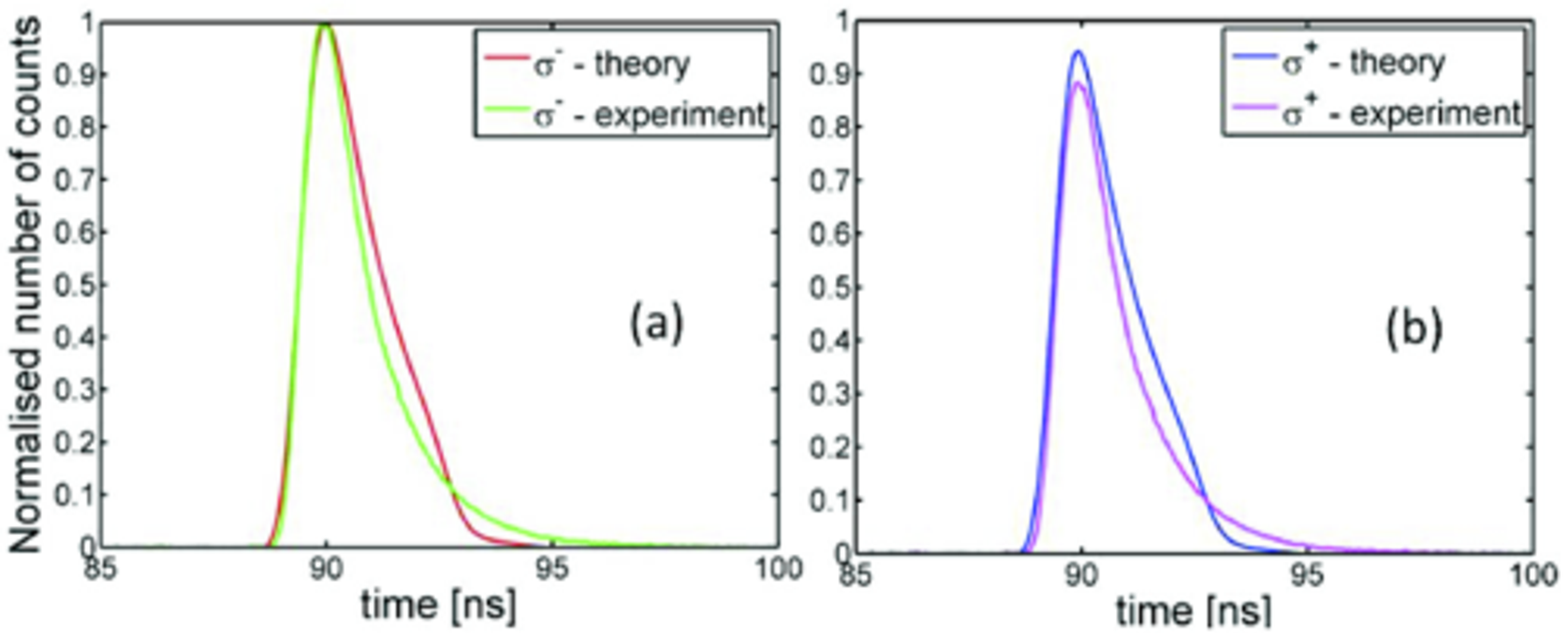}
\vspace{-10pt}
\caption[Fig 10] {\label{fig:Fig 10}(Color online) Comparison between simulated TRPL trace (after convolution with detector response function) and experimentally detected TRPL trace for (a) $\sigma^-$ excitation and (b) $\sigma^+$ excitation.}\label{fig:theory-experiment comparison}
\end{figure*}

The convolved theoretically calculated polarised TRPL traces are plotted along with the experimental data for both pulse helicities in Fig.~\ref{fig:theory-experiment comparison} for the parameters listed in Table I.
%%%%%%%%%%%%%%%%%%%%%%%%%%%%%%%%%%%%%%%%%%%%%%%%%%%%%%%%%%%%%%%%%%%%%%%%%%%%%%%
%%%%%%%%%%%%%%%%%%%%%%%%%%%%%%%%%%%%%%%%%%%%%%%%%%%%%%%%%%%%%%%%%%%%%%%%%%%%%%%%%%%%%
The theoretical results agree remarkably well with the experiment. The rising slope and the decay immediately after the peak are almost perfectly reproduced. The small discrepancy at longer times can be attributed to several factors: there is no way, in the experiment, to determine precisely which hot trion state is resonantly excited, so the assumed $\lambda_{res}$ is only approximate; $\Delta_{s-p}$ is inferred from the resonant PL spectra, which is only a rough estimate, as the excited state can be viewed as an envelope of a number of spectral lines; the inhomogeneous broadening is not taken into account. The lower peak height for $\sigma^+$ excitation can be attributed to the estimated value for the spin flip-flop coupling rate and requires further investigation. This is, however, beyond the scope of this work. The theory correctly predicts circular dichroism in the polarised TRPL emission for both pulse helicities, shown Fig.~\ref{fig:dichroism_comparison}.
%%%%%%%%%%%%%%%%%%%%%%%%%%%%%%%%%%%%%%%%%%%%%%%%%%%%%%%%%%%%%%%%%%%%%%%%%%%%%%%
\begin{figure}
\vspace{-10pt}
%\resizebox{\columnwidth}{!}{\includegraphics{Fig11.eps}}
\includegraphics[height=5 cm, width=6 cm]{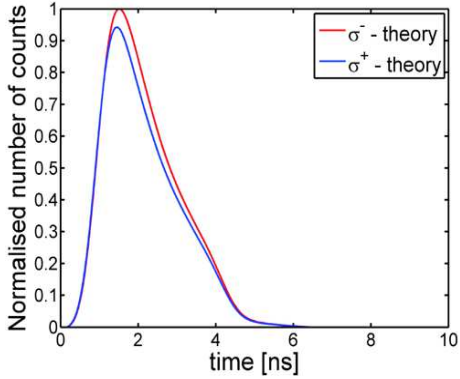}
\vspace{-10pt}
\caption[Fig 11] {\label{fig:dichroism_comparison}(Color online) Theoretically computed polarised TRPL traces for $\sigma^-$ and $\sigma^+$ excitation; simulation parameters from Table I, exhibiting circular dichroism.}\label{fig:dichroism_comparison}
\end{figure}
%%%%%%%%%%%%%%%%%%%%%%%%%%%%%%%%%%%%%%%%%%%%%%%%%%%%%%%%%%%%%%%%%%%%%%%%%%%%%%%%%%%%%

\section{Conclusions}
The proposed new theory and model explain quantitatively the origin of the spin-filtering effect. The observed enhanced time-resolved circular dichroism in the excited trion state emission is shown to emerge from the dynamical spin flip-flop coupling between the two degenerate sets of excited charged excitonic states. Remarkably good agreement between theory and experiment is achieved, thereby yielding largely unknown intra and inter-shell spin-relaxation timescales. The advantageous to this effect increased optical spin injection efficiency under quasi-resonant excitation into QD hot trion states, compared to the ground trion singlet, opens up new avenues for high-fidelity optical spin manipulation. The model represents a tool for investigation and identification of an optimum pulse power for achieving maximum degree of spin polarisation, thus maximizing the efficiency of the optical spin orientation. This enables high-fidelity quantum computing gate operations. The proposed formalism can be extended to describe the spin dynamics in multiply-charged QDs for which further enhancement of the circular dichroism is reported \cite{Taylor}.
\begin{acknowledgments}
The author is indebted to M. Taylor. P. Spencer and E. Clarke for stimulating discussions and making their experimental data available. Funding through EPSRC grant EP/H000488/1 is gratefully acknowledged.
\end{acknowledgments}

\appendix
%\appendixpage
\begin{subappendices}
\section{Generators of SU(6) Lie algebra and pseudospin equations}
\label{sec:1}
For an N-level system there are $N^2-1$ generators,$\lambda_j$, satisfying the commutation relations:
\begin{equation}
\label{eq:commutation relations}
[\hat \lambda _j ,\hat \lambda _k ] = 2i\,f_{jkl} \hat \lambda_l
\end{equation}
where $f_{jkl}$ is the fully antisymmetric tensor of the structure constants of SU(N) Lie algebra. $f$ guarantees constant length of the real state vector, $|\textbf{S}|=const$ \cite{Hioe&Eberly}. For the 6-level system considered, there are 35 lambda-generators. A possible choice satisfying the othogonality relations:
\begin{equation}
Tr\left( {\hat \lambda _j .\hat \lambda _k } \right) = 2\delta _{jk}
\end{equation}
can be constructed from the transition-projection operators:
\begin{equation}
\hat P_{mn}  = \left| m \right\rangle \left\langle n \right|
\end{equation}
where $\left| m \right\rangle$ and $\left| n \right\rangle$ are eigenstates of the system Hamiltonian, using the following definitions:
\begin{equation}
\begin{array}{l}
 \hat u_{jk}  = \hat P_{jk}  + \hat P_{kj}, \,\,\,\,\,\,\,\,\,\,\,\,\,\,\,\,\,\,\,\,\,\,\,\,\,\,\,\,\,\,\,\,\,\,\,1 \le j < k \le N \\
 \hat v_{jk}  =  - i\left( {\hat P_{jk}  - \hat P_{kj} } \right) \\
 \hat w_l  =  - \left[ {\frac{2}{{l\left( {l + 1} \right)}}} \right]^{1/2} \left( {\hat P_{11}  + ... + \hat P_{ll}  - l\hat P_{l + 1,l + 1} } \right),  1 \le l \le N \\
 \end{array}
\end{equation}
with $N=6$ and $l=1,2,..,5$. The vector $\mathbb{\hat \lambda}$, defined as an ordered array of the elements of $\hat u$, $\hat v$ and $\hat w$:
\begin{equation}
{\bf \hat \lambda } = \left( {\hat u_{12} ,...,\hat v_{12} ,...,\hat w_1 ,...,\hat w_{N - 1} } \right)
\end{equation}
has operator components, $\hat \lambda_j$, that satisfy the commutation relations Eq.(\ref{eq:commutation relations}) and are traceless. The latter represent the generators of the SU(6) group.

The fully anstisymmetric tensor, given by:
\begin{equation}
f_{ijk}  = \frac{1}{4}i\left( {Tr\left( {\hat \lambda _i .\hat \lambda _k .\hat \lambda _j } \right) - Tr\left( {\hat \lambda _i .\hat \lambda _j .\hat \lambda _k } \right)} \right)
\end{equation}
has 750 non-vanishing components, and the torque vector is given by:
\begin{equation}
\label{eq:torque vector}{\bf \gamma }_j \left( t \right) = \frac{1}{\hbar }Tr\left( {\hat H(t) \hat
%\mathord{\buildrel{\lower3pt\hbox{$\scriptscriptstyle\frown$}}
\lambda_j } \right)
\end{equation}
where the time-dependent Hamiltonian, $\hat H$, is given by Eq. (1). For $\sigma^-$ excitation Eq. (\ref{eq:torque vector}) gives:
\setlength{\mathindent}{0 cm}
\begin{widetext}
\begin{equation}
\begin{array}{l}
 {\bf \gamma } = (0,\,0,\,0,\,0, - \Omega _x, 0,\,0,\,0,\,0,\,0,\,0,\,0,\,0,\,0,\,0,\,0,\,0,\,0,\,0, - \Omega _y, 0,\,0,\,0,\,0,\,0,\,0,\,0,\,0,\,0,\,0, \\
 \\
  - \Delta _{hh}  - \Delta _{sp}  + \omega _0, \,\frac{{ - 2\Delta _{eh}  - \Delta _{hh}  + \Delta _{sp}  + \omega _0 }}{{\sqrt 3 }},\,\frac{{\Delta _{eh}  - \Delta _{hh}  + \Delta _{sp}  + \omega _0 }}{{\sqrt 6 }},\,\frac{{5\Delta _{eh}  - \Delta _{hh}  + \Delta _{sp}  + \omega _0 }}{{\sqrt {10} }},\,\frac{{4\Delta _{hh}  + \Delta _{sp}  + \omega _0 }}{{\sqrt {15} }}) \\
\end{array}
\end{equation}
\end{widetext}
Substituting the torque vector and the non-vanishing tensor components into the equation of motion Eq.(2) for the real pseudospin vector, $\textbf{S}$:
\begin{equation}
\frac{{\partial S_i }}{{\partial t}} = \sum\limits_{j = 1}^{35} {\sum\limits_{k = 1}^{35} {f_{ijk} \gamma _j S_k } } ,\,\,\,\,i = 1,2,...,35
\end{equation}

a set of 35 equations for the real state vector components is obtained, which can be written in a matrix form:
\begin{equation}
{\bf \dot S} = {\bf \hat MS}
\end{equation}
where the dot denotes time derivative and the matrix $\hat M$ is a $35 \times 35$ matrix, depending on the energy level splittings and the Rabi frequencies, $\Omega_x$ and $\Omega_y$. Due to the problem symmetry, the pseudospin equations remain valid for $\sigma^+$ excitation, provided that $\Omega_y$ is replaced by $-\Omega_y$.

The dissipation in the system, through spin relaxation and decoherence processes, is taken into account within the master pseudospin formalism \cite{Slavcheva} and is explained in the following section.

\section{Longitudinal and transverse spin relaxation matrices}
\label{sec:2}
Consider $\sigma^-$ optical excitation of the $|1\rangle\rightarrow|6\rangle$ transition (Fig. 2, left panel) and subsequent relaxation via all dipole-allowed transitions, with corresponding decay rates, to the ground state with spin-up heavy hole. The population of each level satisfies the following set of master equations for the diagonal density matrix components:

\begin{align}
%\begin{equation}
%\begin{array}{l}
& \scalemath{0.95} {\frac{{\partial \hat \rho _{11} }}{{\partial t}} = \frac{i}{\hbar }\left[ {\hat \rho ,\hat H} \right]_{11}  + \Gamma _{61} \hat \rho _{66}  + \Gamma _{51} \hat \rho _{55}  + \Gamma _{41} \hat \rho _{44}  + \Gamma _{21} \hat \rho _{22}} \nonumber \\
& \scalemath{0.95} {\frac{{\partial \hat \rho _{22} }}{{\partial t}} = \frac{i}{\hbar }\left[ {\hat \rho ,\hat H} \right]_{22}  + \Gamma _{62} \hat \rho _{66}  + \Gamma _{52} \hat \rho _{55}  - \Gamma _{21} \hat \rho _{22}} \nonumber \\
& \scalemath{0.95} {\frac{{\partial \hat \rho _{33} }}{{\partial t}} = \frac{i}{\hbar }\left[ {\hat \rho ,\hat H} \right]_{33}  + \gamma _{43} \hat \rho _{44}  - \gamma _{34} \hat \rho _{33}}  \\
& \scalemath{0.95} {\frac{{\partial \hat \rho _{44} }}{{\partial t}} = \frac{i}{\hbar }\left[ {\hat \rho ,\hat H} \right]_{44}  + \gamma _{54} \hat \rho _{55}  - \left( {\gamma _{45}  + \gamma _{43}  + \Gamma _{41} } \right)\hat \rho _{44}  + \gamma _{34} \hat \rho _{33}} \nonumber \\
& \scalemath{0.95}{\frac{{\partial \hat \rho _{55} }}{{\partial t}} = \frac{i}{\hbar }\left[ {\hat \rho ,\hat H} \right]_{55}  + \Gamma _{65} \hat \rho _{66}  + \gamma _{45} \hat \rho _{44}  - \left( {\gamma _{54}  + \Gamma _{51}  + \Gamma _{52} } \right)\hat \rho _{55}}  \nonumber \\
& \scalemath{0.95} {\frac{{\partial \hat \rho _{66} }}{{\partial t}} = \frac{i}{\hbar }\left[ {\hat \rho ,\hat H} \right]_{66}  - \left( {\Gamma _{61}  + \Gamma _{62}  + \Gamma _{65} } \right)\hat \rho _{66}} \nonumber
%\end{array}
%\end{equation}
\end{align}

where $\Gamma_{ij}$ and $\gamma_{ij}$ are the relaxation rate between levels $i$ and $j$ and I have introduced, for the sake of generality, different rates of spin population transfer, due to e-h anisotropic exchange interaction, between the excited triplet levels (denoted by $\gamma_{AEI}$ in Fig. 3).
For each of the above equations we can introduce a $6\times6$ matrix, $\hat \Gamma_{i}, \,\,\ i=1,2,..,6$, describing the spin relaxation rates as follows:
%\begin{eqnarray}
%\begin{widetext}
\begin{eqnarray*}
\label{eq:Gammai}
%\begin{align*}
\begin{array}{l}
\scalemath{0.6} {\hat \Gamma _1  = \left( {\begin{array}{*{20}c}
   0 & 0 & 0 & 0 & 0 & 0  \\
   0 & {\Gamma _{21} } & 0 & 0 & 0 & 0  \\
   0 & 0 & 0 & 0 & 0 & 0  \\
   0 & 0 & 0 & {\Gamma _{41} } & 0 & 0  \\
   0 & 0 & 0 & 0 & {\Gamma _{51} } & 0  \\
   0 & 0 & 0 & 0 & 0 & {\Gamma _{61} } \\
\end{array}} \right)\,\,\hat \Gamma _2  = \left( {\begin{array}{*{20}c}
   0 & 0 & 0 & 0 & 0 & 0  \\
   0 & { - \Gamma _{21} } & 0 & 0 & 0 & 0  \\
   0 & 0 & 0 & 0 & 0 & 0  \\
   0 & 0 & 0 & 0 & 0 & 0  \\
   0 & 0 & 0 & 0 & {\Gamma _{52} } & 0  \\
   0 & 0 & 0 & 0 & 0 & {\Gamma _{62} }  \\
\end{array}} \right) }

\end{array}
%\end{align*}
\end{eqnarray*}
\begin{eqnarray}
%\begin{align}
%\label{eqn:Gammai}
\begin{array}{l}
\scalemath{0.6} {\hat \Gamma _3  = \left( {\begin{array}{*{20}c}
   0 & 0 & 0 & 0 & 0 & 0  \\
   0 & 0 & 0 & 0 & 0 & 0  \\
   0 & 0 & { - \gamma _{34} } & 0 & 0 & 0  \\
   0 & 0 & 0 & {\gamma _{43} } & 0 & 0  \\
   0 & 0 & 0 & 0 & 0 & 0  \\
   0 & 0 & 0 & 0 & 0 & 0  \\
\end{array}} \right) \,\,
\hat \Gamma _4  = \left( {\begin{array}{*{20}c}
   0 & 0 & 0 & 0 & 0 & 0  \\
   0 & 0 & 0 & 0 & 0 & 0  \\
   0 & 0 & {\gamma _{34} } & 0 & 0 & 0  \\
   0 & 0 & 0 & { - \left( {\Gamma _{41}  + \gamma _{43}  + \gamma _{45} } \right)} & 0 & 0  \\
   0 & 0 & 0 & 0 & {\gamma _{54} } & 0  \\
   0 & 0 & 0 & 0 & 0 & 0 \notag\\
\end{array}} \right) }
\end{array}
%\end{align}
\end{eqnarray}
\begin{eqnarray}
\begin{aligned}
%\begin{equation}
\begin{array}{l}
\scalemath{0.6} {\hat \Gamma _5  = \left( {\begin{array}{*{20}c}
   0 & 0 & 0 & 0 & 0 & 0  \\
   0 & 0 & 0 & 0 & 0 & 0  \\
   0 & 0 & 0 & 0 & 0 & 0  \\
   0 & 0 & 0 & {\gamma _{45} } & 0 & 0  \\
   0 & 0 & 0 & 0 & { - \left( {\Gamma _{51}  + \Gamma _{52}  + \gamma _{54} } \right)} & 0  \\
   0 & 0 & 0 & 0 & 0 & {\Gamma _{65} }  \\
\end{array}} \right) \,\,
\hat \Gamma _6  = \left( {\begin{array}{*{20}c}
   0 & 0 & 0 & 0 & 0 & 0  \\
   0 & 0 & 0 & 0 & 0 & 0  \\
   0 & 0 & 0 & 0 & 0 & 0  \\
   0 & 0 & 0 & 0 & 0 & 0  \\
   0 & 0 & 0 & 0 & 0 & 0  \\
   0 & 0 & 0 & 0 & 0 & { - \left( {\Gamma _{61}  + \Gamma _{62}  + \Gamma _{65} } \right)}  \\
\end{array}} \right)} \text{ \label{eqn:Gammai}} \\
\end{array}
\end{aligned}
%\end{equation}
\end{eqnarray}
%\end{widetext}

Then we can define $\hat \sigma=diag(Tr(\hat \Gamma_i \hat \rho))$, thus isolating the longitudinal spin relaxation from the transverse one and write the equation of motion for the density matrix:
\vspace{-20pt}
\begin{equation}
\label{eq:Liouville}\frac{{\partial \hat \rho }}{{\partial t}} = \frac{i}{\hbar }\left[ {\hat \rho ,\hat H} \right] + \hat \sigma  - \hat \Gamma _t \hat \rho
\end{equation}
The transverse spin relaxation (decoherence) matrix, $\hat \Gamma_t$, is obtained by writing down master equations for the off-diagonal density matrix components, taking into account the spin decoherence mechanisms relevant for each level (see Fig. 3, double-headed blue arrows coupling the two 6-level systems): e.g. for the ground level -- hole-spin decoherence rate $\gamma_h$, due to dipole-dipole hyperfine interaction with the lattice ion spins, for the levels with unpaired electron spin -- $\gamma_e$, electron spin decoherence via hyperfine interaction, and for the excited triplet levels -- $\gamma_{ff}$, spin flip-flop  processes, whereby the electron and the hole simultaneously flip their spin. As a result, the transverse spin decoherence matrix reads:
\begin{equation}
\label{eq:Gamma_t}\hat \Gamma _t  = \left( {\begin{array}{*{20}c}
   0 & {\gamma _e } & {\gamma _h } & {\gamma _h } & {\gamma _h } & {\gamma _h }  \\
   {\gamma _e } & 0 & {\gamma _e } & {\gamma _e } & {\gamma _e } & {\gamma _e }  \\
   {\gamma _{ff} } & {\gamma _{ff} } & 0 & {\gamma _{ff} } & {\gamma _{ff} } & {\gamma _{ff} }  \\
   {\gamma _{ff} } & {\gamma _{ff} } & {\gamma _{ff} } & 0 & {\gamma _{ff} } & {\gamma _{ff} }  \\
   {\gamma _e } & {\gamma _e } & {\gamma _e } & {\gamma _e } & 0 & {\gamma _e }  \\
   {\gamma _e } & {\gamma _e } & {\gamma _e } & {\gamma _e } & {\gamma _e } & 0  \\
\end{array}} \right)
\end{equation}

The density matrix, $\hat \rho$ can be expressed in terms of the lambda-generators of SU(N) Lie algebra, where $N=6$ is the number of discrete levels of the quantum system, as follows:
\begin{equation}
\label{eq:rho}\hat \rho \left( t \right) = \frac{1}{N}\hat I + \frac{1}{2}\sum\limits_{j = 1}^{N^2  - 1} {S_j \left( t \right)} \hat \lambda _j
\end{equation}

where $S_j$ are the real state vector components and $\hat I$ is the unit matrix.
The elements of the diagonal $\hat \sigma$ matrix in terms of the real state vector components can be obtained by taking the trace of the product of $\hat \rho$ from Eq.(\ref{eq:rho}) with the $\hat \Gamma_i$ matrices from Eq.(\ref{eq:Gammai}):
\setlength{\mathindent}{0pt}
%\begin{align}
%\begin{widetext}
\begin{equation}
\begin{array}{l}
\label{eq:sigma matrix}
\scalemath{0.75} {
\hat \sigma _{11}  = \frac{1}{{60}}[\left( {10 + 30S_{31}  - 10\sqrt 3 S_{32}  - 5\sqrt 6 S_{33}  - 3\sqrt {10} S_{34}  - 2\sqrt {15} S_{35} } \right)\Gamma _{21}}\\
 \scalemath{0.75} {\hspace{20pt} + 15\sqrt 6 S_{33} \Gamma _{41}  - 3\sqrt {10} \left( {\Gamma _{41}  - 4\Gamma _{51} } \right)S_{34}  - 2\sqrt {15} \left( {\Gamma _{41}  + \Gamma _{51}  - 5\Gamma _{61} } \right)S_{35} } \\
 \scalemath{0.75}{\hspace{20pt}+ 10\left( {\Gamma _{41}  + \Gamma _{51}  + \Gamma _{61} } \right)] } \\
 \\
\scalemath{0.75}{ \hat \sigma _{22}  = \frac{1}{{60}}[\left( { - 10 - 30S_{31}  + 10\sqrt 3 S_{32}  + 5\sqrt 6 S_{33}  + 3\sqrt {10} S_{34}  + 2\sqrt {15} S_{35} } \right)\Gamma _{21} } \\
\scalemath{0.75}{ \hspace{20pt} + 2\left( {5 + 6\sqrt {10} S_{34}  - \sqrt {15} S_{35} } \right)\Gamma _{52}}
\scalemath{0.75}{ \hspace{20pt} + 10\left( {1 + \sqrt {15} S_{35} } \right)\Gamma _{62} ] }\\
 \\
 \scalemath{0.75}{\hat \sigma _{33}  = \frac{1}{{60}}[\left( { - 10 - 20\sqrt 3 S_{32}  + 5\sqrt 6 S_{33}  + 3\sqrt {10} S_{34}  + 2\sqrt {15} S_{35} } \right)\gamma _{34}} \\
 \scalemath{0.75}{\hspace{20pt}+ \left( {10 + 15\sqrt 6 S_{33}  - 3\sqrt {10} S_{34}  - 2\sqrt {15} S_{35} } \right)\gamma _{43} ]}\\
 \\
\scalemath{0.75}{\hat \sigma _{44}  = \frac{1}{{60}}[\left( {10 + 20\sqrt 3 S_{32}  - 5\sqrt 6 S_{33}  - 3\sqrt {10} S_{34}  - 2\sqrt {15} S_{35} } \right)\gamma _{34}  }\\
 \scalemath{0.75}{\hspace{20pt}+ 2\left( {5 + 6\sqrt {10} S_{34}  - \sqrt {15} S_{35} } \right)\gamma _{54} } \\
 \scalemath{0.75}{\hspace{20pt}- \left( {10 + 15\sqrt 6 S_{33}  - 3\sqrt {10} S_{34}  - 2\sqrt {15} S_{35} } \right)\left( {\gamma _{43}  + \gamma _{45}  + \Gamma _{41} } \right)] } \\
\\
\scalemath{0.75}{ \hat \sigma _{55}  = \frac{1}{{60}}\left( {10 + 15\sqrt 6 S_{33}  - 3\sqrt {10} S_{34}  - 2\sqrt {15} S_{35} } \right)\gamma _{45}}\\
\scalemath{0.75}{ \hspace{20pt}- 2\left( {5 + 6\sqrt {10} S_{34}  - \sqrt {15} S_{35} } \right)\left( {\gamma _{54}  + \Gamma _{51}  + \Gamma _{52} } \right) } \\
\scalemath{0.75}{ \hspace{20pt}+ 10\left( {1 + \sqrt {15} S_{35} } \right)\Gamma _{65} ] }\\
 \\
\scalemath{0.75}{ \hat \sigma _{66}  =  - \frac{1}{6}\left( {1 + \sqrt {15} S_{35} } \right)\left( {\Gamma _{61}  + \Gamma _{62}  + \Gamma _{65} } \right)}
\end{array}
\end{equation}
%\end{align}
%\end{widetext}

Substituting $\hat \sigma$ into the second line of Eq.(2), we can find expressions for the relaxation times of the spin population terms towards their equilibrium values, which for $\sigma^-$ excitation read:

\begin{align}
%\begin{array}{l}
& T_{31}  = \frac{2}{{\Gamma _{21} }} \nonumber \\
& T_{32}  = \frac{3}{{\gamma _{34} }} \nonumber \\
& T_{33}  = \frac{{24}}{{4\gamma _{34}  + 12\gamma _{43}  + 9\gamma _{45}  + 12\Gamma _{41} }}\\
& T_{34}  = \frac{8}{{\gamma _{45}  + 4\left( {\gamma _{54}  + \Gamma _{51}  + \Gamma _{52} } \right)}} \nonumber \\
&T_{35}  = \frac{2}{{\Gamma _{61}  + \Gamma _{62}  + \Gamma _{65}}} \nonumber
%\end{array}
\end{align}

The transverse spin decoherence times, $T_j, j=1,2,...,30$ in the first line of Eq. (2), are given in terms of the spin decoherence rates by:
\begin{align}
%\begin{equation}
%\begin{array}{l}
\label{eq:transverse_decoherence}
& \scalemath{0.9}{T_1  = T_9  = T_{15}  = T_{16}  = T_{23}  = T_{24}  = T_{30}  = \frac{1}{{\gamma _e }} }\nonumber \\
& \scalemath{0.9}{T_2  = T_3  = T_{17}  = T_{18}  = \frac{2}{{\gamma _h \left( {1 + \frac{{\gamma _{ff} }}{{\gamma _h }}} \right)}}} \nonumber \\
& \scalemath{0.9}{T_4  = T_5  = T_{19}  = T_{20}  = \frac{2}{{\gamma _h \left( {1 + \frac{{\gamma _e }}{{\gamma _h }}} \right)}}} \\
& \scalemath{0.9}{T_6  = T_7  = T_8  = T_{21}  = T_{22}  = \frac{2}{{\gamma _e \left( {1 + \frac{{\gamma _{ff} }}{{\gamma _e }}} \right)}} }\nonumber \\
& \scalemath{0.9}{T_{10}  = T_{25}  = \frac{1}{{\gamma _{ff} }} }\nonumber \\
& \scalemath{0.9}{T_{11}  = T_{12}  = T_{13}  = T_{14}  = T_{26}  = T_{27}  = T_{28}  = T_{29}  = \frac{2}{{\gamma _{ff} \left( {1 + \frac{{\gamma _e }}{{\gamma _{ff} }}} \right)}}} \nonumber
%\end{array}
%\end{equation}
\end{align}

In the case of a $\sigma^+$ excitation, the energy-level diagram is more complex, as additional relaxation channels are dynamically switched on (Fig. 4). The population of each level satisfies a different set of master equations for the diagonal density matrix components:
\begin{align}
%\begin{equation}
%\begin{array}{l}
& \scalemath{0.8}{\frac{{\partial \rho _{11} }}{{\partial t}} = \frac{i}{\hbar }\left[ {\hat \rho ,\hat H} \right]_{11}  + \Gamma _{61} \rho _{66}  + \Gamma _{51} \rho _{55}  + \Gamma _{41} \rho _{44}  + \Gamma _{21} \rho _{22}} \nonumber \\
& \scalemath{0.8}{\frac{{\partial \rho _{22} }}{{\partial t}} = \frac{i}{\hbar }\left[ {\hat \rho ,\hat H} \right]_{22}  + \Gamma _{62} \rho _{66}  + \Gamma _{52} \rho _{55}  + \Gamma _{42} \rho _{44}  + \Gamma _{32} \rho _{33}  - \Gamma _{21} \rho _{22} }\nonumber \\
& \scalemath{0.8}{\frac{{\partial \rho _{33} }}{{\partial t}} = \frac{i}{\hbar }\left[ {\hat \rho ,\hat H} \right]_{33}  + \Gamma _{63} \rho _{66}  + \Gamma _{43} \rho _{44}  - \left( {\gamma _{34}  + \Gamma _{32} } \right)\rho _{33}}  \\
& \scalemath{0.8}{\frac{{\partial \rho _{44} }}{{\partial t}} = \frac{i}{\hbar }\left[ {\hat \rho ,\hat H} \right]_{44}  + \Gamma _{64} \rho _{66}  + \gamma _{54} \rho _{55}  - \left( {\gamma _{45}  + \gamma _{43}  + \Gamma _{41}  + \Gamma _{42} } \right)\rho _{44}  + \gamma _{34} \rho _{33}} \nonumber \\
& \scalemath{0.8}{\frac{{\partial \rho _{55} }}{{\partial t}} = \frac{i}{\hbar }\left[ {\hat \rho ,\hat H} \right]_{55}  + \Gamma _{65} \rho _{66}  - \left( {\gamma _{54}  + \Gamma _{51}  + \Gamma _{52} } \right)\rho _{55}  + \gamma _{45} \rho _{44}} \nonumber \\
&\scalemath{0.8}{\frac{{\partial \rho _{66} }}{{\partial t}} = \frac{i}{\hbar }\left[ {\hat \rho ,\hat H} \right]_{66}  - \left( {\Gamma _{65}  + \Gamma _{64}  + \Gamma _{63}  + \Gamma _{62}  + \Gamma _{61} } \right)\rho _{66}} \nonumber
%\end{array}
%\end{equation}
\end{align}
We can introduce, similar to the $\sigma^-$ case, $6\times6$ matrices, $\hat \Gamma_{i}, \,\,\ i=1,2,..,6$, describing the spin relaxation rates as follows:
\begin{eqnarray}
\begin{array}{l}
\label{eq:Gamma_i_sigma_plus} \scalemath{0.6}{ \hat \Gamma _1  = \left( {\begin{array}{*{20}c}
   0 & 0 & 0 & 0 & 0 & 0  \\
   0 & {\Gamma _{21} } & 0 & 0 & 0 & 0  \\
   0 & 0 & 0 & 0 & 0 & 0  \\
   0 & 0 & 0 & {\Gamma _{41} } & 0 & 0  \\
   0 & 0 & 0 & 0 & {\Gamma _{51} } & 0  \\
   0 & 0 & 0 & 0 & 0 & {\Gamma _{61} }  \\
\end{array}} \right)}
\scalemath{0.6}{\,\,\hat \Gamma _2  = \left( {\begin{array}{*{20}c}
   0 & 0 & 0 & 0 & 0 & 0  \\
   0 & { - \Gamma _{21} } & 0 & 0 & 0 & 0  \\
   0 & 0 & {\Gamma _{32} } & 0 & 0 & 0  \\
   0 & 0 & 0 & {\Gamma _{42}}& 0 & 0  \\
   0 & 0 & 0 & 0 & {\Gamma _{52} } & 0  \\
   0 & 0 & 0 & 0 & 0 & {\Gamma _{62} }  \\
\end{array}} \right)} \hspace{40pt} \text{\label{eq:Gamma_i_sigma_plus}}
\end{array}
\end{eqnarray}
%\end{eqnarray}
\\
\\
\begin{eqnarray*}
\begin{array}{l}
\scalemath{0.55}{\hat \Gamma _3  = \left( {\begin{array}{*{20}c}
   0 & 0 & 0 & 0 & 0 & 0  \\
   0 & 0 & 0 & 0 & 0 & 0  \\
   0 & 0 & { - (\gamma _{34}+\Gamma_{32}) } & 0 & 0 & 0  \\
   0 & 0 & 0 & {\gamma _{43} } & 0 & 0  \\
   0 & 0 & 0 & 0 & 0 & 0  \\
   0 & 0 & 0 & 0 & 0 & {\Gamma _{63}}  \\
\end{array}} \right) }
\scalemath{0.55}{\,\,\hat \Gamma _4  = \left( {\begin{array}{*{20}c}
   0 & 0 & 0 & 0 & 0 & 0  \\
   0 & 0 & 0 & 0 & 0 & 0  \\
   0 & 0 & {\gamma _{34} } & 0 & 0 & 0  \\
   0 & 0 & 0 & { - \left( {\gamma _{45}  + \gamma _{43}+\Gamma _{41}+\Gamma _{42} } \right)} & 0 & 0  \\
   0 & 0 & 0 & 0 & {\gamma _{54} } & 0  \\
   0 & 0 & 0 & 0 & 0 & \Gamma _{64}  \\
\end{array}} \right)}
\end{array}
\end{eqnarray*}
\\
\begin{eqnarray*}
\begin{aligned}
\scalemath{0.55}{\hat \Gamma _5  = \left( {\begin{array}{*{20}c}
   0 & 0 & 0 & 0 & 0 & 0  \\
   0 & 0 & 0 & 0 & 0 & 0  \\
   0 & 0 & 0 & 0 & 0 & 0  \\
   0 & 0 & 0 & {\gamma _{45} } & 0 & 0  \\
   0 & 0 & 0 & 0 & { - \left( {\Gamma _{51}  + \Gamma _{52}  + \gamma _{54} } \right)} & 0  \\
   0 & 0 & 0 & 0 & 0 & {\Gamma _{65} }  \\
\end{array}} \right)}
\scalemath{0.55}{\,\,\hat \Gamma _6  = \left( {\begin{array}{*{20}c}
   0 & 0 & 0 & 0 & 0 & 0  \\
   0 & 0 & 0 & 0 & 0 & 0  \\
   0 & 0 & 0 & 0 & 0 & 0  \\
   0 & 0 & 0 & 0 & 0 & 0  \\
   0 & 0 & 0 & 0 & 0 & 0  \\
   0 & 0 & 0 & 0 & 0 & { - \left( {\Gamma _{61}  + \Gamma _{62}  +\Gamma _{63}+\Gamma _{64}+\Gamma _{65} } \right)}  \\
\end{array}} \right)}  \\
%\end{array}
\end{aligned}
\end{eqnarray*}

The longitudinal spin relaxation can be incorporated in a diagonal matrix, $\hat \sigma$, whose components are calculated as the $Tr(\hat \Gamma_i \hat \rho)$ and subsequently expressed in terms of the real state vector components, $S_j, j=31,32,...,35$, responsible for the spin population difference between the pairs of levels. The longitudinal spin relaxation times in Eq.(2) are given by:
\begin{align}
%\begin{equation}
%\begin{array}{l}
& T_{31}  = \frac{2}{{\Gamma _{21} }} \nonumber \\
& T_{32}  = \frac{6}{{3\Gamma _{32}  - 2\gamma _{34} }} \nonumber \\
& T_{33}  = \frac{{24}}{{4\gamma _{34}  + 12\gamma _{43}  + 9\gamma _{45}  + 12\left( {\Gamma _{41}  + \Gamma _{42} } \right)}}\\
& T_{34}  = \frac{8}{{\gamma _{45}  + 4\gamma _{54}  + 4\left( {\Gamma _{51}  + \Gamma _{52} } \right)}} \nonumber \\
& T_{35}  = \frac{2}{{\Gamma _{61}  + \Gamma _{62}  + \Gamma _{63}  + \Gamma _{64}  + \Gamma _{65} }} \nonumber
%\end{array}
%\end{equation}
\end{align}

The transverse decoherence times are given by Eq.(\ref{eq:transverse_decoherence}).

\section{Relationship between the macroscopic polarisation and real state vector components}
\label{sec:3}
The macroscopic medium polarisation is given by the expectation value of the dipole moment operator, $e\langle\hat Q\rangle$, where $\hat Q$ is the local displacement operator.
\begin{equation}
P =  - N_a e\left\langle {\hat Q} \right\rangle  =  - eN_a Tr\left( {\hat \rho .\hat Q} \right)
\end{equation}
where $N_a$ is the density of the polarisable atoms (QDs) in the medium.
Since the electric field vector of a circularly polarised pulse is in a plane perpendicular to the propagation (growth) z-direction:
\begin{equation}
\label{eq:electric field}{\bf E} = E_x {\bf e}_{\bf x}  + E_y {\bf e}_{\bf y}
\end{equation}
the dipole moment induced in the medium will be along $E_x$ and $E_y$ electric field vector components, therefore the local displacement operator can be decomposed in two components, along x-- and y-- axis, selecting appropriate lambda-generators in such a manner that they produced a coherent, circularly polarised excitation term in the dipole interaction Hamiltonian:
\begin{equation*}
\scalemath{0.9} {\hat Q = q_0 \left\{ {\left( {\begin{array}{*{20}c}
   0 & 0 & 0 & 0 & 0 & 1  \\
   0 & 0 & 0 & 0 & 0 & 0  \\
   0 & 0 & 0 & 0 & 0 & 0  \\
   0 & 0 & 0 & 0 & 0 & 0  \\
   0 & 0 & 0 & 0 & 0 & 0  \\
   1 & 0 & 0 & 0 & 0 & 0  \\
\end{array}} \right){\bf e}_{\bf x}  + \left( {\begin{array}{*{20}c}
   0 & 0 & 0 & 0 & 0 & { - i}  \\
   0 & 0 & 0 & 0 & 0 & 0  \\
   0 & 0 & 0 & 0 & 0 & 0  \\
   0 & 0 & 0 & 0 & 0 & 0  \\
   0 & 0 & 0 & 0 & 0 & 0  \\
   i & 0 & 0 & 0 & 0 & 0  \\
\end{array}} \right){\bf e}_{\bf y} } \right\}}
\end{equation*}
\begin{equation}
\label{eq:displacement}
= q_0 \left( {\hat \lambda _5 {\bf e}_{\bf x}  + \hat \lambda _{20} {\bf e}_y } \right)
\end{equation}
where $\textbf{e}_x$ and $\textbf{e}_y$ are the unit vectors along $x-$ and $y-$ axes, $q_0$ is the typical distance between the charges in a dipole and the dipole moment is given by $\wp=eq_0$.
With this choice of lambda-generators and substituting $\textbf{E}$ and $\hat Q$ from Eq.(\ref{eq:electric field}) and (\ref{eq:displacement}), the dipole-coupling interaction Hamiltonian acquires its form from Eq.(1):
\begin{equation*}
\hat H_{{\mathop{\rm int}} } \left( t \right) =  - e{\bf E}.{\bf \hat Q} =  - eq_0 \left( {E_x \hat \lambda _5  + E_y \hat \lambda _{20} } \right)
\end{equation*}
\begin{equation}
= \hbar \left( {\begin{array}{*{20}c}
   0 & 0 & 0 & 0 & 0 & { - \frac{1}{2}\left( {\Omega _x  \mp i\Omega _y } \right)}  \\
   0 & 0 & 0 & 0 & 0 & 0  \\
   0 & 0 & 0 & 0 & 0 & 0  \\
   0 & 0 & 0 & 0 & 0 & 0  \\
   0 & 0 & 0 & 0 & 0 & 0  \\
   { - \frac{1}{2}\left( {\Omega _x  \pm i\Omega _y } \right)} & 0 & 0 & 0 & 0 & 0  \\
\end{array}} \right)
\end{equation}
where the upper (lower) sign denotes $\sigma^-$ ($\sigma^+$) excitation and we have defined the complex Rabi frequency components $\Omega _x  = \wp \frac{{E_x }}{\hbar }$, $\Omega _y  = \wp \frac{{E_y }}{\hbar }$.

Taking the trace of the product of the density matrix operator in terms of the real state vector, given by Eq.(B5) and the local displacement operator Eq. (C3), one can obtain Eq.(4).
%%%%%%%%%%%%%%%%%%%%%%%%%%%%%%%%%%%%%%%%%%%%%%%%%%%%%%%%%%%%%%%%%%%%%%%%%%%%%%%%%
\begin{figure*}
%\vspace{20pt}
%\includegraphics[height=12 cm, width=16 cm]{Fig12.eps}
%\resizebox{\columnwidth}{!}{\includegraphics{Fig12.eps}}
\includegraphics[scale=0.7]{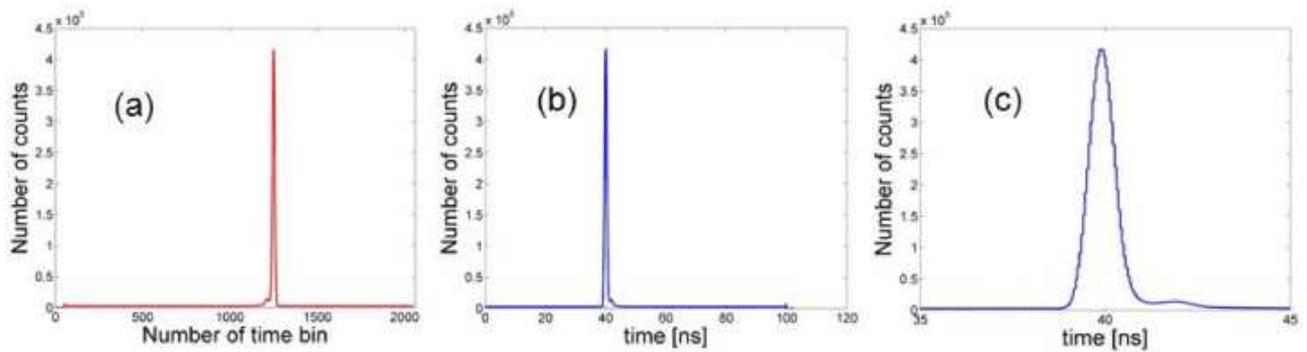}
%\vspace{-10pt}
\caption[Fig 12] {\label{fig:detector_response_function}(Color online)(a) InP/InGaAsP PMT detector response function, as given by the specifications; (b) Converted in the time domain with sampling rate $\Delta t_1=0.05\,\mathrm{ns}$; (c) Discretised with step size of the simulation sampling rate, $\Delta t_2=3.33\times10^{-4} \,\mathrm{fs}$.}
\end{figure*}
%%%%%%%%%%%%%%%%%%%%%%%%%%%%%%%%%%%%%%%%%%%%%%%%%%%%%%%%%%%%%%%%%%%%%%%%%%%%%%%%%%%
\begin{figure*}
%\vspace{20pt}
%\includegraphics[height=12 cm, width=16 cm]{Fig13.eps}
%\resizebox{\columnwidth}{!}{\includegraphics{Fig13.eps}}
\includegraphics[scale=0.7]{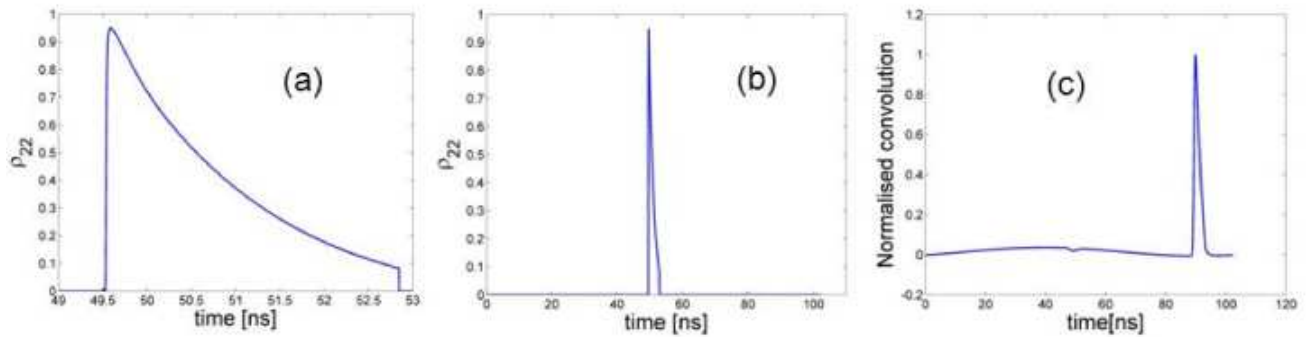}
%\vspace{-10pt}
\caption[Fig 13] {\label{fig:convolution}(Color online) (a) Simulation data for $\rho_{22}$ versus time;  (b) Simulation data padded with zeros to a length in time matching the one of the detector response function; (c) Convolution of the simulation data in (b) with the discretised detector response function from Fig. \ref{fig:detector_response_function}(c). Note that as a result of the convolution, the signal is translated along the time axis.}
\end{figure*}

\section{Convolution of the simulation data with detector response function}
\label{sec:4}
The detector response function of the InP/InGaAsP PMT detector, as given by its specifications, shown in Fig. \ref{fig:detector_response_function} (a), is converted in the time domain, Fig. \ref{fig:detector_response_function} (b). Since the detector sampling rate $\Delta t_1=0.05 \,\,\mathrm{ns}$ is much coarser than the simulations sampling rate of $\Delta t_2=3.33 \times 10^{-4} \,\,\mathrm{fs}$, we add $n=[\Delta t_1/\Delta t_2]$ points within each bin and thus discretise it, Fig. \ref{fig:detector_response_function}(c).

In order to calculate the convolution of the simulation data, $g$, with the detector response function, $h$:
\begin{equation}
g*h = \int\limits_{ - \infty }^\infty  {g\left( \tau  \right)} h\left( {t - \tau } \right)d\tau
\end{equation}
we take advantage of the Convolution Theorem:
\begin{equation}
F(g*h) \Leftrightarrow G(f)H(f)
\end{equation}
where G and H are the Fourier-transforms of the convolved functions, respectively $g$, and $h$, and calculate inverse Fourier transform of the product of Fourier-transforms above. The simulation data is padded with zeros to match the length (in number of points) of the detector response function and is convolved with the latter. The functions before and after the convolution are shown in Fig. \ref{fig:convolution}.

\end{subappendices}
%\end{appendix}

% *********************** REFERENCES ********************

%\newpage

% *********************** FIGURES ********************

\newpage

\end{document}